\documentclass[%
 aip,
 jmp,%
 amsmath,amssymb,
 reprint, numerical%
]{revtex4-2}

\usepackage{graphicx}
\usepackage{dcolumn}
\usepackage{bm}
\usepackage{natbib}
\usepackage{amsmath}

\begin{document}

\preprint{AIP/123-QED}

\title{Optical atomic clocks on all-optical atom chips}

\author{Yuri B. Ovchinnikov}
\affiliation{National Physical Laboratory, Hampton Road, Teddington TW11 0LW, United Kingdom}

\date{\today}

\begin{abstract}
The feasibility and potential performance of optical clocks based on all-optical atom chips using neutral strontium atoms have been investigated. The key benefits of all-optical atom chips are their ability to trap alkali-earth atoms, and how easily these atoms can be manipulated and examined. The planar configuration of the chip enables ring photonic resonators to be incorporated into its structure, making it a promising candidate for active optical atomic clocks. Using the 'magic wavelengths' of the trapping optical fields (813 nm and 390 nm for strontium atoms) minimises frequency shifts in the intercombination clock transition caused by variations in optical field intensity. However, this frequency shift can be compensated for using optical AC Stark shifts, which are generated by the trapping optical fields and have frequencies below the 'magic frequencies'. Furthermore, it has been demonstrated that atom chip clocks can operate using solely repulsive, blue-detuned evanescent light fields when the attraction of the trapped atoms towards the photonic waveguide is realised through van der Waals forces alone. According to estimates, this could result in the short-term fractional frequency stability of the strontium atom-chip optical clocks being limited to $~10^{-16}$ per second, with an accuracy of about $5\times10^{-15}$ for fractional intensity stability in the trapping light fields of $0.001$.
\end{abstract}

\maketitle

\section{Introduction}
The idea of incorporating atomic clocks into atom chips is gaining popularity. There are several reasons for this phenomenon. First, atom chip atomic clocks are expected to be small, lightweight, and power-efficient. Second, the same atom chip can incorporate various components of atomic clocks, such as lasers, atomic sources, atomic traps, and frequency micro-combs. In addition, the same atom chip can be used to incorporate other devices. For example, a quantum navigator chip could incorporate an atomic clock, quantum inertial sensors (e.g., accelerometers, gyroscopes, and gravimeters), quantum memory, and quantum artificial intelligence (AI).

The most advanced approach to atom chips \cite{reichel2011atom} involves trapping neutral atoms in magnetic field gradients generated by current-carrying wires integrated into the chip substrate. Microwave rubidium clocks based on standard magnetic atom chips are being developed \cite{szmuk2015stability,dupont2025noise}. These clocks demonstrated long-term frequency stability of the order of $5\times10^{-13}$. This is primarily limited by fluctuations in the trapping magnetic field.

The other class of miniature microwave atomic clocks is called chip-scale atomic clocks (CSAC) \cite{knappe2004microfabricated}, which are based on miniature atomic vapor cells. These clocks are based on coherent population trapping (CPT) in rubidium or cesium atoms and have an accuracy between $10^{-10}$ and $10^{-12}$. Technically, these are not atomic chips because the atoms being interrogated are not confined to the substrate of the chip. 

Another type of miniature integrated atomic clock is based on a two-photon $5S_{1/2}\rightarrow5D_{5/2}$ optical electronic transition in rubidium \cite{newman2021high}. This optical frequency standard demonstrated long-term frequency stability of approximately $10^{-14}$. The main limitation to the short-term frequency stability of such clocks is the relatively large natural linewidth of this optical transition of 670\,kHz, compared to clocks based on forbidden optical transitions. It has recently been proposed that rubidium atoms trapped in two-colour evanescent light waves using silicon nitride strip waveguides \cite{harding2026system} could be used to realise an optical clock based on the $5S_{1/2}\rightarrow5D_{5/2}$ electronic transition. The main problem with such clocks is that the van der Waals interaction of rubidium atoms with the dielectric surface of the photonic waveguide in the $5D_{5/2}$ state is about one order of magnitude larger than for the $5S_{1/2}$ state. This should result in a significant frequency shift of the clock transition and may cause additional non-uniform broadening.

In this paper, we discuss the perspective of optical atomic clocks based on alkali earth atoms trapped in two-color evanescent light fields generated by optical modes propagating in suspended planar photonic waveguides of all-optical atomic chips \cite{ovchinnikov2020towards}.  
It should be noted that it is not possible to trap alkali earth atoms in standard magnetic atomic chips \cite{reichel2011atom} because these atoms have a zero-spin ground state, making them untrappable in magnetic field gradients.

The main advantages of optical lattice atomic clocks \cite{takamoto2005optical}  are the use of ultra-narrow intercombination-line transitions of alkaline-earth atoms and the 'magic' wavelengths of trapping optical lattices. These provide trapping optical potentials for atoms without affecting the frequencies of the corresponding optical-clock transitions. Strontium optical atomic clocks \cite{aeppli2024clock} recently reached an accuracy \cite{jia2026improved} of $9.2 \times 10^{-19}$, making them one of the most accurate atomic clocks. The high quality factor of the clock transitions of alkali-earth atoms provides optical atomic clocks with a very high frequency stability. The strontium optical lattice atomic clock exhibits a short-term frequency stability of $2.2\times10^{-16}$ over a one-second period, as detailed Ref. \cite{nicholson2015systematic}. This stability is primarily constrained by fluctuations in the interrogating laser frequency.

These clocks have become popular because of the experimental convenience of accessing all the laser frequencies necessary for two-stage laser cooling of strontium atoms, trapping them in a 'magic' optical lattice, and subsequently interrogating the optical clock transition. This makes strontium optical lattice clocks one of the leading candidates for the future redefinition of the second in the International System of Units (SI) \cite{dimarcq2024roadmap,fortier2026optical}, which is currently defined via the hyperfine ground microwave transition of cesium atoms. Therefore, developing compact strontium atomic clocks based on atom chips for widespread industrial use is a current issue.

Compared to all-optical atom chips (AOAC), trapping atoms in two-color evanescent light fields formed by the propagating modes of nanofibers is already a well-established experimental technology, as demonstrated in Refs. \cite{nieddu2016optical, solano2017optical}. Recently, strontium atoms were trapped in a nanofiber at 435 and 473\,nm magic wavelengths of the trapping light for the $5^1S_0 \rightarrow 5 ^3P_1,|m|=1$ intercombination line transition, as demonstrated in Ref. \cite{kestler2023state}. However, this transition has a linewidth of approximately 7.4\,kHz, making it less attractive for use in atomic clocks than the true ultra-narrow clock transition of strontium atoms. Another issue with optical nanofibers is their high sensitivity to mechanical vibrations, which results in rapid heating and loss of trapped atoms.  

The remainder of this paper is organized as follows: The second section considers the general design of the AOAC strontium optical clock and describes the modeling of its optical fields. The third section describes the calculation of the trapping potentials of the clocks, including both the optical dipole and van der Waals potentials. It also considers the frequency shifts of the strontium clock transition of the trapped atoms in the presence of trapping potentials. The fourth section analyzes the different possible regimes of the clocks and their potential performance.     

\section{Design of AOAC clock}

\subsection{General structure of the clock}

The concept behind AOAC optical clocks involves trapping alkali-earth atoms within a surface optical lattice formed by evanescent light fields generated by the two-color modes of suspended photonic rib waveguides \cite{ovchinnikov2020towards}.

At first we assume that the wavelengths of all the waveguide trapping optical modes are magic. For strontium atoms, this means that the attractive red-frequency-detuned mode has a wavelength of $\lambda_r=813$\,nm \cite{takamoto2005optical} and the repulsive blue-frequency-detuned mode has a wavelength of $\lambda_b=390$\,nm \cite{takamoto2009prospects}. The strongest strontium optical electronic transition, $5^1S_0\rightarrow5^1P_1$, has a wavelength of 461\,nm and is primarily responsible for trapping atoms in the ground state, $5^1S_0$.
To enhance the atomic trapping efficiency, it is assumed that the suspended photonic waveguide supports two-mode propagation (TE$_{00}$ and TE$_{01}$) of 390\,nm light and single-mode propagation (TE$_{00}$) of 813\,nm light. This makes trapping atoms in two-color evanescent light fields more efficient \cite{ovchinnikov2025two}. In order to form the surface optical lattice, which is necessary for the efficient operation of optical atomic clocks, it is supposed that the red-frequency-detuned light at 813\,nm includes two counter-propagating TE$_{00}$ modes of equal intensity. The corresponding spatial distribution of the surface optical red-frequency-detuned lattice is illustrated in Fig.\,1. 

\begin{figure}
	\includegraphics[width=10.5 cm]{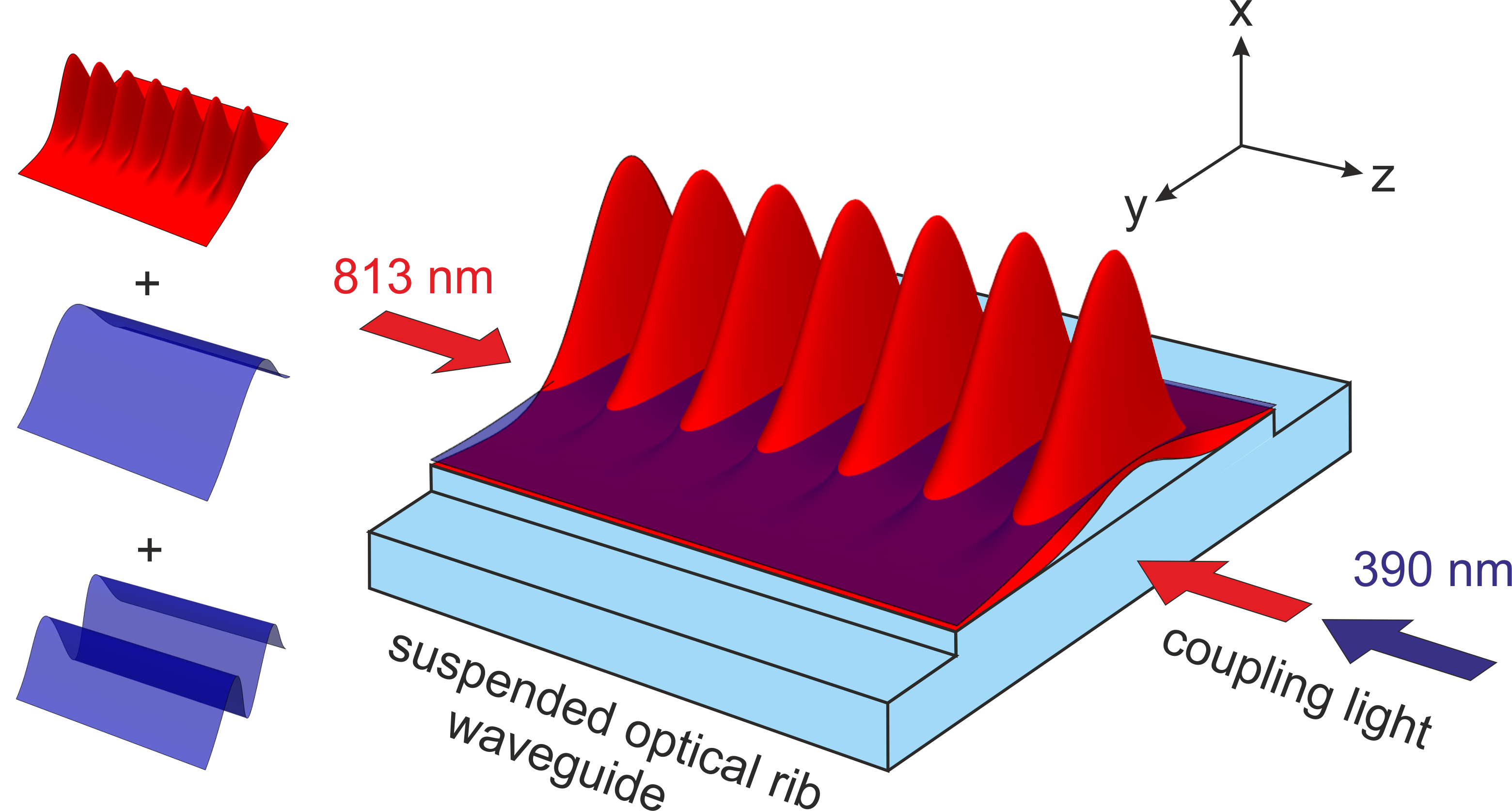}
	\caption{The figure shows a spatial distribution of light intensity in the $y0z$ plane at the surface of suspended photonic rib waveguide, which consists of two counterpropagating TE$_{00}$ modes at 813\,nm and a composition of TE$_{00}$ and TE$_{01}$ modes at 390\,nm. \label{fig1}}
\end{figure}  

Another way to trap atoms in the waveguide is to use a superposition of a repulsive, blue-frequency-detuned surface optical lattice and an attractive van der Waals force. This approach has recently been successfully demonstrated for optical nanofibers \cite{pennetta2025hybrid}. Such a blue-detuned surface optical lattice can be formed by the superposition of the TE$_{00}$ and TE$_{01}$ modes, with the  TE$_{01}$ mode having a larger amplitude than the TE$_{00}$ mode. It is also assumed that the TE$_{00}$ mode comprises two counter-propagating components with different amplitudes. This makes it possible to adjust the relative amplitude of the optical-lattice component. This creates minima in the optical dipole potential in the $y0z$ plane, as shown in Fig.\,2. Combined with a spatially uniform attractive van der Waals potential, this optical dipole potential can provide stable trapping of atoms in the blue-detuned optical surface lattice.   

\begin{figure}
	\includegraphics[width=10.5 cm]{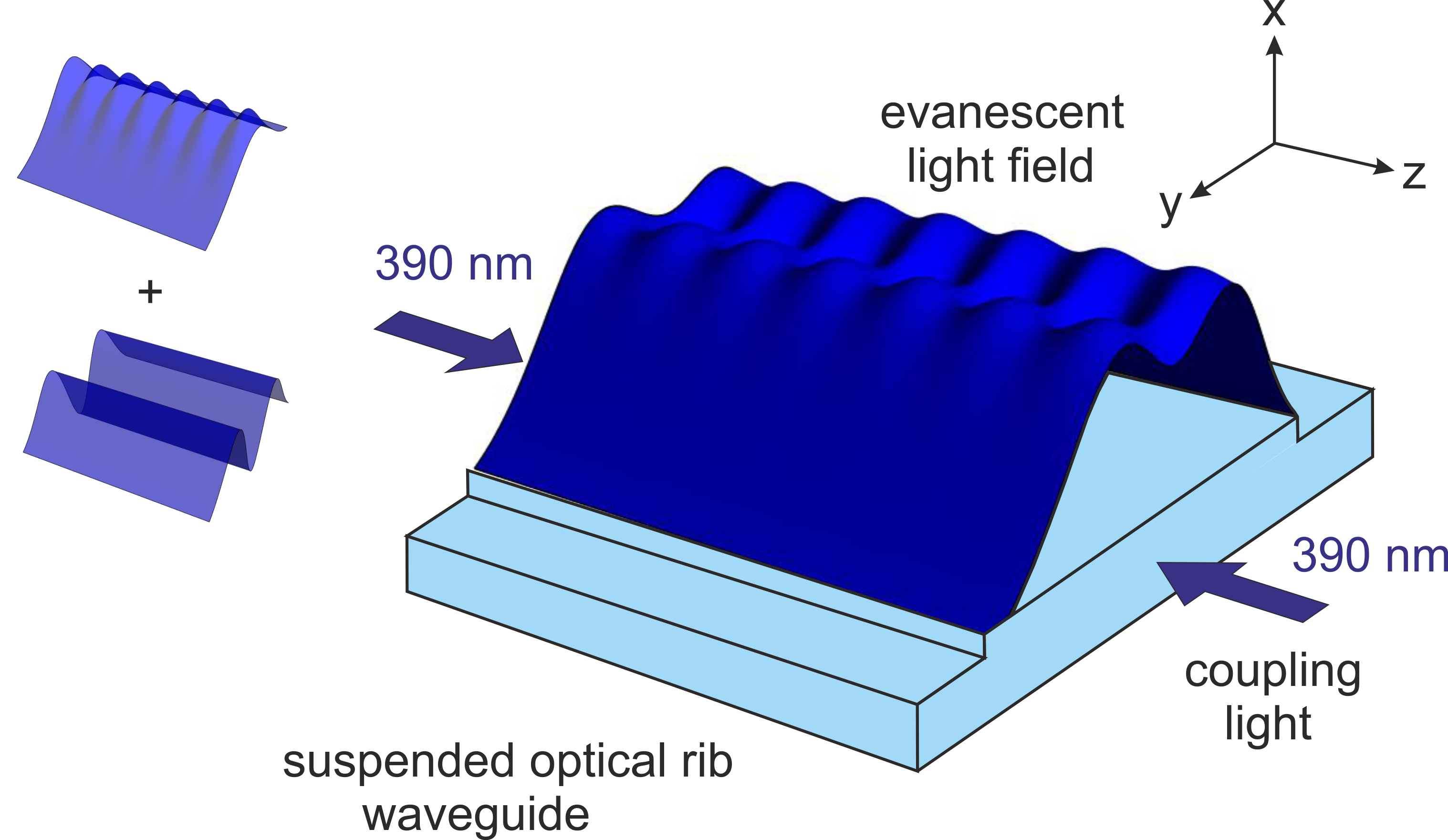}
	\caption{The figure shows a spatial distribution of light intensity in the $y0x$ plane at the surface of suspended photonic rib waveguide. \label{fig2}}
\end{figure}  

According to numerical simulations \cite{Hammer2025solver}, a suspended optical rib waveguide made of silica with a membrane thickness of $h_m=150$\,nm, rib height of $h_r=30$\,nm \cite{ovchinnikov2025two} and rib width of $w=1.4$\,$\mu$m supports the propagation of a single TE$_{00}$ mode at 813\,nm with an effective refractive index of the waveguide $n_{eff}(\lambda_r,\text{TE}_{00})=1.1524$ and a propagation constant of $\beta(\lambda_r,\text{TE}_{00})=8.9061$\,$\mu$m$^{-1}$. In contrast, 390\,nm light includes two modes: TE$_{00}$ with $n_{eff}(\lambda_b,\text{TE}_{00})=1.3187$ and $\beta(\lambda_b,\text{TE}_{00})=21.2460$\,$\mu$m$^{-1}$ and TE$_{01}$ with $n_{eff}(\lambda_b,\text{TE}_{01})=1.3059$ and $\beta(\lambda_b,\text{TE}_{01})=21.0383$ $\mu$m$^{-1}$. The corresponding penetration depths of the evanescent waves of these modes in vacuum are $d_r=213.8$\,nm and $d_b=71.4$\,nm. The same waveguide supports the propagation of a single TE$_{00}$ mode at $\lambda_c=698$\,nm with effective index $n_{eff}(\lambda_c, TE_{00})=1.1856$ and propagation constant $\beta(\lambda_c, TE_{00})=10.6722$\,$\mu$m$^{-1}$, which can be used to interrogate the $5^1S_0\rightarrow 5^3P_0$ clock transition of strontium atoms. 

Now, let's address the important issue of interference between the TE$_{00}$ and TE$_{01}$ modes of the waveguide at 390\,nm. When the phase difference between the TE$_{00}$ and TE$_{01}$ modes at each point along the waveguide at 390 nm is fixed, interference between the modes produces a zigzag pattern of varying intensity along the waveguide, with a period of $\Lambda=\lambda_b/(n_{eff}(\lambda_b,\text{TE}_{00})-n_{eff}(\lambda_b,\text{TE}_{01}))=30.5$\,$\mu$m. In that case, the two modes only exhibit a symmetric intensity distribution when their relative phases are equal to $\pi/2 \pm n \pi$, where $n=0,1,2...$.

Several methods can be employed to avoid or suppress interference between the TE$_{00}$ and TE$_{01}$ modes. These methods either eliminate coherence between the two modes or cause the corresponding interference pattern to move rapidly along the waveguide. Shifting the interference pattern rapidly along the waveguide creates a consistent, time-averaged potential for the trapped atoms, essentially eliminating the time-averaged interference term between the two modes with opposite transverse symmetries. The first method involves exciting the TE$_{00}$ and TE$_{01}$ modes of the waveguide independently and selectively using two lasers with nearly identical wavelengths. Such a selective excitation of the TE$_{00}$ and TE$_{01}$ modes in a planar photonic waveguide can be achieved using a tapered photonic waveguide and a directional coupler, as demonstrated in Ref.\cite{dorin2014two}. Random and rapid fluctuations in the relative phases of the light produced by the independent lasers would lead to an absence of coherence between the two modes in this case. The second method could involve rapidly modulating the phase of the light to the mode that has been selectively excited. This would result in a rapid shift of the corresponding interference pattern along the waveguide. If the relative phase between the modes changes by $\pi$ (or more) much faster than the period of the atoms' lateral oscillations in the trapping potential (typically 0.1\,ms), the atoms will only experience the time-averaged potential. An alternative to modulating the relative phase of the two modes' optical fields could be changing the relative frequencies of the two modes. Estimates show that shifting the frequency of one mode by $\Delta \omega/2\pi=10$\,MHz should cause the corresponding interference pattern to travel along the waveguide at a speed of $\Delta \omega/\Delta \beta \approx 300$\,m/s. This would result in the interference pattern shifting by one full period in just 0.1\,$\mu$s. Another method involves using an integrated electro-optic modulator \cite{sinatkas2021electro} to modulate the phases of the propagating modes directly within the photonic waveguide. Due to the different propagation constants and spatial distributions of the TE$_{00}$ and TE$_{01}$ modes, such a modulator will cause a differential phase shift between the modes. This could be used to rapidly move the corresponding interference pattern. The advantage of this method is that independent excitation and control of the two modes is not required.

Therefore, in the subsequent sections the problem of eliminating interference between the TE$_{00}$ and TE$_{01}$ modes of the waveguide at 390 \, nm has been resolved using one of the methods given above.

\section{Energies of clock states and trapping potentials}

The total external potential experienced by the strontium atoms in their $5^1S_0$ and $5^3P_0$ clock states, $U_{tot}^{(i)}\left(\mathbf{r}\right)$, where the indices $i=s,p$ correspond to the ground and excited states of the clock transition, formed by the evanescent light fields of the propagating modes of the suspended photonic rib waveguide, including the van der Waals interaction between the trapped atoms and the waveguide's dielectric surface, is calculated as follows:

\begin{equation} \label{EQ1}
\resizebox{0.92 \textwidth} {!}
{$U_{tot}^{(i)}\left(\mathbf{r}\right) = U_{r,\text{TE}_{00}}^{(i)}\left(y,z\right) e^{-2x/d_r}+\left(U_{b,\text{TE}_{00}}^{(i)}\left(y,z\right) + U_{b,\text{TE}_{01}}^{(i)}\left(y,z\right)\right) e^{-2x/d_b}-\frac{\epsilon-1}{\epsilon+1}\frac{C_3^{(i)}}{x^3}$},
\end{equation}

The first three terms correspond to the optical dipole potentials of the evanescent wave fields of the TE$_{00}$ red-frequency-detuned and two TE$_{00}$ and TE$_{01}$ blue-frequency-detuned optical modes of the waveguide. The last term represents the interaction potential between the strontium atoms and the dielectric surface of the SiO$_2$ rib-waveguide. The van der Waals coefficients are $C_3^{(s)} = 2.190 \times 10^{-48}$\,Jm$^3$ and $C_3^{(p)} = 2.352 \times 10^{-48}$\,Jm$^3\;$  \cite{derevianko2010electric}, while $\epsilon=3.9$ is the electric permittivity of silicon dioxide. The origin of the x-axis is located on the surface of the dielectric waveguide. 
	
The lateral distributions of the optical dipole potentials, $U_{r,TE_{00}}^{(i)}$, $U_{b,TE_{00}}^{(i)}$ and $U_{b,TE_{01}}^{(i)}$, along the surface of the waveguide rib were calculated using the modelled intensity distributions of light for the corresponding optical modes, according to the formula $U\left(\mathbf{r},\lambda\right)^{(i)}=-\alpha(\lambda)^{(i)} I(\mathbf{r})/(2\epsilon_0 c)$, where $\mathbf{r}$ is a spatial coordinate, $\epsilon_0
$ is the vacuum permittivity, $c$ is the velocity of light and $\alpha(\lambda)^{(i)}$ is the dynamic polarizability of the strontium atoms \cite{safronova2015extracting}. The polarizability of strontium atoms for the $5^1S_0$ and $5^3P_0$ clock states, as a function of wavelength, were calculated according to the publications \cite{safronova2015extracting, lisdat2021blackbody}, and are shown in Fig.\,3. The green dashed line shows the polarizability of the $5^1S_0$ state and the blue dashed line shows the polarizability of the $5^3P_0$ state. The red solid line shows the difference between the polarizabilities of the $5^3P_0$ and $5^1S_0$ states. The zero values of the differential polarizability of the clock states correspond to the magic wavelengths of the strontium atoms. The red point corresponds to the 813\,nm magic wavelength, and the blue dot corresponds to the 390\,nm wavelength. Below this wavelength, blue-detuned optical fields form repulsive optical dipole potentials, whereas above this wavelength, red-detuned attractive optical dipole potentials are formed. 
	
	\begin{figure}
	\includegraphics[width=12 cm]{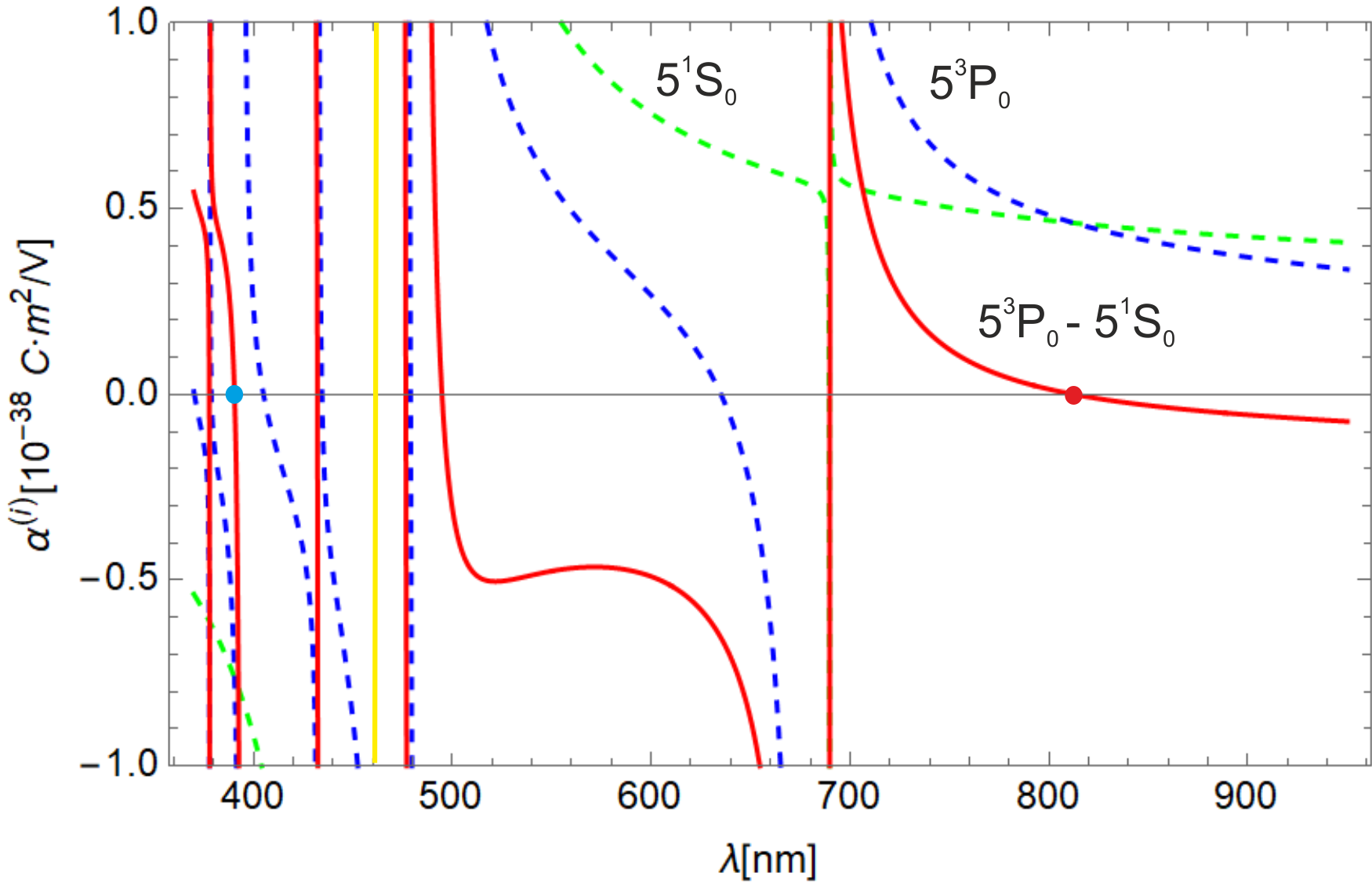}
	\caption{The frequency-dependent polarizabilities of the Sr $5^1S_0$ (green dashed line) and the $5^3P_0$ (blue dashed line) clock states, as well as their differential polarizability (red solid line). The yellow vertical line shows the 461\,nm wavelength of the main and strongest $5^1S_0 \rightarrow 5^1P_1$ transition.\label{fig3}}
	\end{figure} 

There are several different configurations of trapping light fields, each with its own designation. One configuration could be responsible for the efficient loading of the AOAC waveguide with cold atoms. The other configurations could be used for the subsequent operation of the AOAC clocks with minimal perturbations in the clock transition.
	
\subsection{Loading potential of the waveguide} 
	
The AOAC waveguide is loaded with cold atoms from another conventional source, such as a magneto-optical trap, optical molasses, optical dipole trap, or a combination of these, in a process known as the loading regime. To achieve this, the trapping potential depth must be maximized, given the available laser power. The propagating modes of the waveguide include a red-detuned  TE$_{00}$ at the magic wavelength of 813\,nm, as well as two co-propagating blue-detuned TE$_{00}$ and TE$_{01}$ modes at the magic wavelength of 390\,nm. The dynamic polarizability of the strontium clock states at 813\,nm is taken to be $4.614 \times 10^{-39}$\,Cm$^2$/V and at 390\,nm to be $-7.564 \times 10^{-39}$\,Cm$^2$/V.
The cross-section of the trapping potential in the $x0y$-plane at maximum surface light intensities $I_{r,TE_{00}}(0,0)=8\times10^{9}$\,W/m$^2$ and $I_{b,TE_{00}}(0,0)=4\times10^{10}$\,W/m$^2$, $I_{b,TE_{01}}(0,y_m)=2\times10^{10}$\,W/m$^2$ is shown in Fig.\,4.  This corresponds to a total laser power of 3.8\,mW for the TE$_{00}$ mode at 813\,nm, 15.9\,mW for the TE$_{00}$ mode at 390\,nm, and 8.9\,mW for the TE$_{01}$ mode at 390\,nm.
		
	\begin{figure}
	\includegraphics[width=10.5 cm]{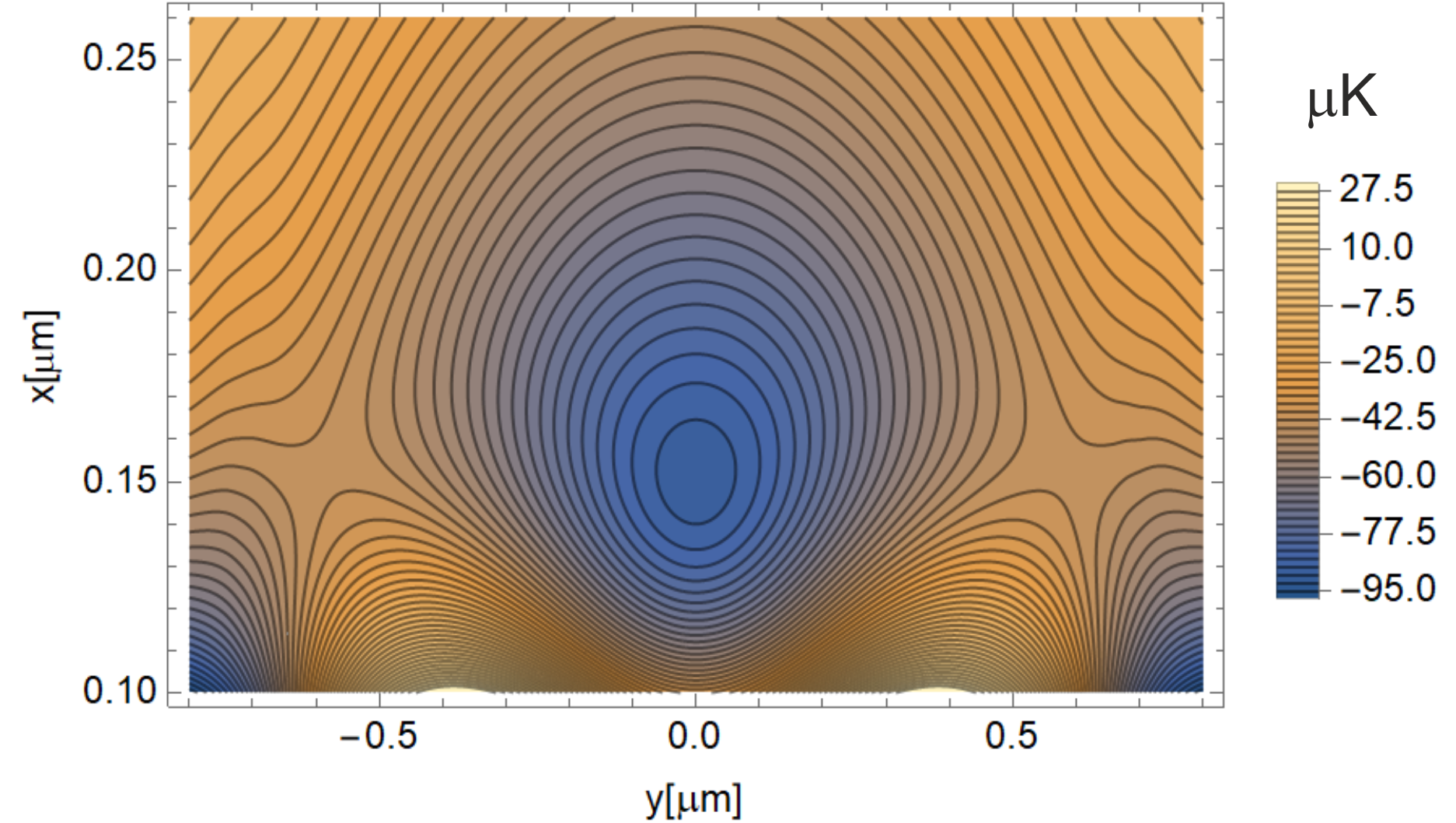}
	\caption{Cross section of total loading potential for one TE$_{00}$ mode of a suspended optical rib waveguide at $\lambda_r=813$ nm and two modes TE$_{00}$ and TE$_{01}$ at $\lambda_b=390$\,nm. \label{fig4}}
	\end{figure}  
			
The minimum of the trap is located at $y=0$, $x_0=152$\,nm, and has a depth of 89\,$\mu$K. The potential depth in the lateral direction is determined by the potential depths at the saddle points located at $x_s=153$\,nm, $y_s=\pm 0.58$\,$\mu$m. The total potential at these points was 43.9\,$\mu$K, which limited the lateral depth of the trapping potential to 45.1\,$\mu$K. Given that the typical temperature of atoms cooled in a MOT using the narrow $5^1S_0-5^3P_1$ transition is approximately 12\,$\mu$K, the trapping depth of the waveguide should suffice.
	
Switching to the clock regime for the loading regime of the trapping potential implies an adiabatic reduction in the powers of the trapping modes to their final target values.
	
\subsection{Clock potential for 390 nm and 813 nm magic trapping light}
	
First, we consider the configuration of the surface optical lattice, as shown in Fig.\,1. This consists of a red-detuned evanescent-field lattice formed by two counter-propagating TE$_{00}$ modes of equal amplitude at the magic wavelength of $\lambda_r=813$\,nm, as well as two co-propagating blue-detuned TE$_{00}$ and TE$_{01}$ modes at the magic wavelength of $\lambda_b=390$\,nm.
The shift in the optical clock transition at the magic wavelengths of the trapping optical fields is expected to be solely due to the difference in van der Waals attraction between the two internal clock states of the atom. To minimize the impact of the van der Waals shift on the clock transition, the atoms should be shifted as far as possible from the waveguide surface. This requires a maximum of 390\,nm repulsive light intensity and a minimum of 813\,nm attractive light intensity. In this scenario, the maximum distance that the atoms can be trapped from the surface ($x_0$) is limited by the maximum light intensity that can be achieved in the waveguide and the minimum potential depth of the trap. This enables stable trapping of the atoms for the required interrogation time of the clock state to be achieved.
The corresponding spatial distribution of the trapping potential in the $y0z$ plane, at $x=x0=257.4$\,nm for strontium atoms in their ground state, $5^1S_0$, at maximum surface light intensities of $I_{r,TE_{00}}^{\pm}(0,0)=1\times10^{9}$\,W/m$^2$ and $I_{b,TE_{00}}(0,0)=4\times10^{10}$\,W/m$^2$, $I_{b,TE_{01}}(0,y_m)=2\times10^{10}$\,W/m$^2$ is shown in Fig.\,5. The spatial distribution of 813\,nm light along the $z$-axis at $x=y=0$ is described by $I_{r,TE_{00}}(0,0,z)=4 I_{r,TE_{00}}^{\pm}(0,0) \cos^2{(2 \pi n_{eff} z/\lambda_{r})}$, where $I_{r,TE_{00}}^{\pm}(0,0)$ are the intensities of the two counter-propagating TE$_{00}$ modes at 813\,nm, and $\lambda_r/n_{eff}=705.5$\,nm is their effective wavelength.

	\begin{figure}
	\includegraphics[width=10.5 cm]{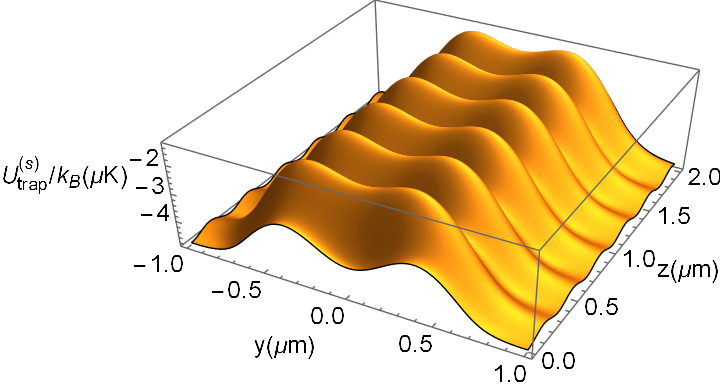}
	\caption{Spatial distribution of the total trapping potential in $y0z$ plane, at $x=x0=257.4$\,nm for strontium atoms in their ground state $5^1S_0$. \label{fig5}}
	\end{figure}  
		
Figure\,6 shows the cross-section of the potential in the $x0z$ plane at $y=0$, which represents a surface optical lattice. The period of the red lattice corresponding to this is equal to $\lambda_r/(2n_{eff})=352.8$\,nm. The amplitude of this surface optical lattice is $\Delta U_0^{lat}=9.2 E_{rec}^{lat}$, where $E_{rec}^{lat}=h^2 n_{eff}^2/(2M \lambda_r^2)$ is the one-photon recoil energy of the lattice, and $M$ is the mass of the strontium atom. This is a relatively deep optical lattice in which the atoms in the lowest energy state are mostly localized within the individual sites of the lattice and can be described by Wannier functions. According to the Bose-Hubbard (for bosons) and Fermi-Hubbard (for fermions) theories, the estimated hopping rate of atoms between the neighboring sites of the optical lattice should be approximately $J/h \sim 120$\,Hz, where $J \approx (4/\sqrt{\pi})E_{rec}^{lat} (\Delta U_0^{lat}/E_{rec}^{lat})^{3/4} \exp[-2 \sqrt{\Delta U_0^{lat}/E_{rec}^{lat}}]$ is the tunneling matrix element. This hopping rate  corresponds to low collision rates between atoms, particularly when there is only one atom per site in the optical lattice. The Lamb-Dicke parameter of the lattice is given by the expression $\eta=k_{clock}/\sqrt{\hbar/(2M\Omega_z)}=0.41$, where $k_{clock}=2 \pi/\lambda_{clock}$, with $\lambda_{clock}=698$\,nm being the wavelength of strontium clock transition and $f_z=\Omega_z/(2\pi)=27.5$\,kHz is the oscillation frequency of the strontium atom near the bottom of the optical lattice. This value of the Lamb-Dicke parameter implies a relatively strong localization of the atoms within the lattice sites and significant suppression of the Doppler broadening of the clock transition related to atomic motion.
		
	\begin{figure}
	\includegraphics[width=10.5 cm]{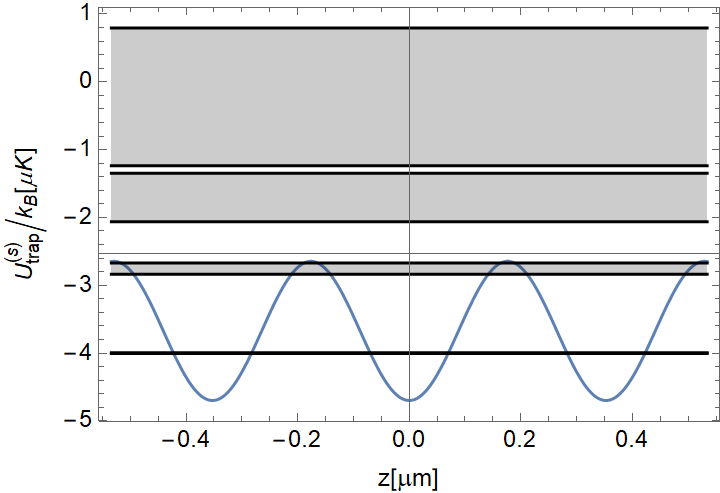}
	\caption{Potential of a surface optical lattice formed by two counter-propagating modes at $813$\,nm, and the corresponding allowed energy bands of the atoms. \label{fig6}}
	\end{figure}  
	
Figure\,7 shows the cross-section of the potential in the $x0y$ plane at $z$ = 0, where the optical lattice has a potential minimum. The lateral depth of the potential $\Delta U_{tot}^{(s)}/k_B=0.76$\,$\mu$K, where $k_B$ is a Boltzmann constant, contains only two bound energy states. The trapped atoms can tunnel along the $y$-axis through the two potential barriers outside the trap. The potential is characterized by the harmonic oscillations frequency near its bottom, which is $f_y=(1/(2\pi))\sqrt{d^2U_{tot}^{(s)}(x=x_0,y=0)/d y^2/M}=7.6$\,kHz. Considering this oscillation frequency and the calculated tunneling probabilities of atoms through the side potential barriers, which are $P_{tun}(n=0)=5.4 \times 10^{-5}$ and $P_{tun}(n=1)=2.0 \times 10^{-2}$, the tunnelling rates of atoms in these two bound states can be estimated as $\Gamma_{tun}(n=0)=0.2$\,s$^{-1}$ and $\Gamma_{tun}(n=1)=212$\,s$^{-1}$.  
		
	\begin{figure}
	\includegraphics[width=10.5 cm]{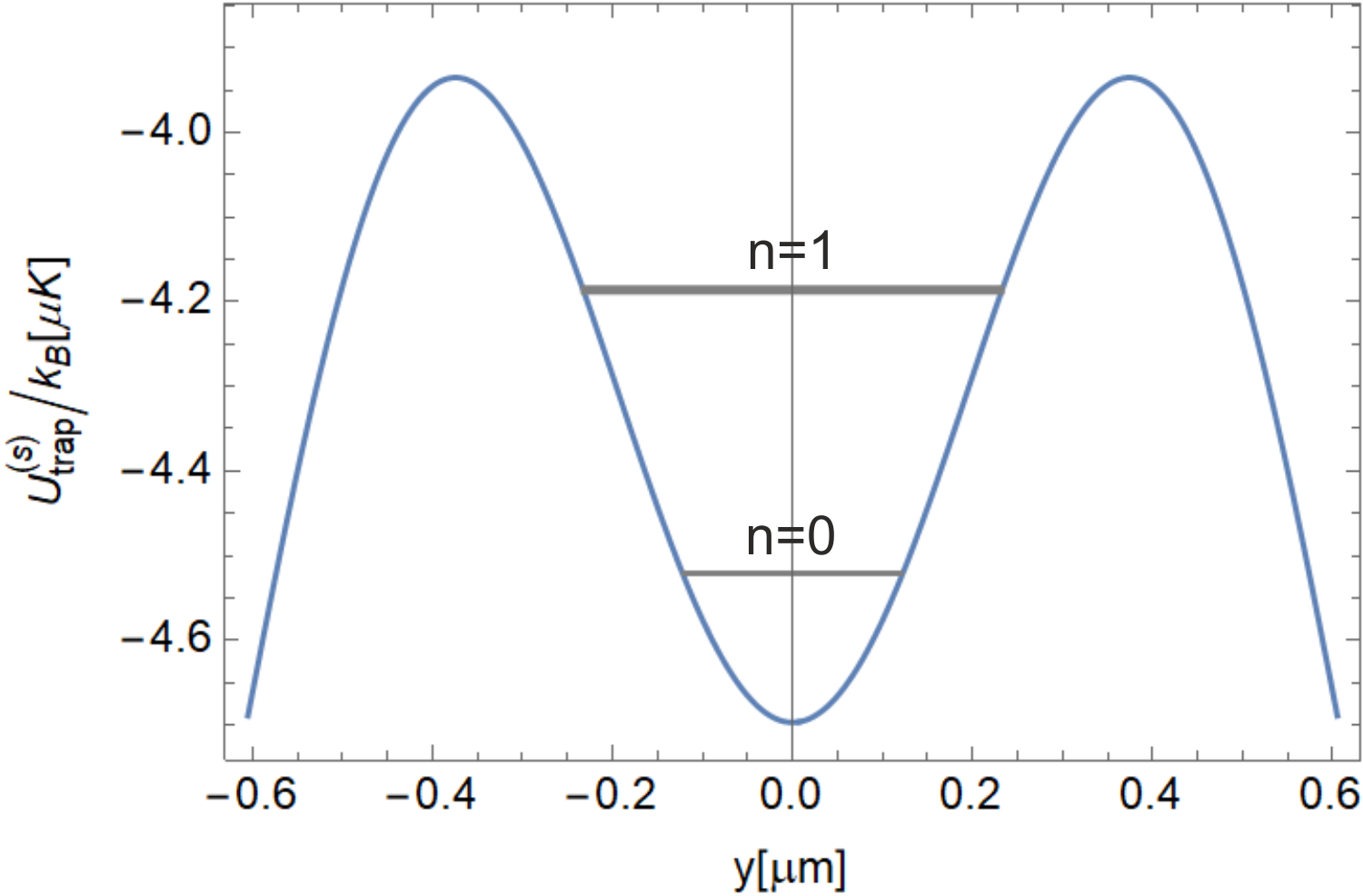}
	\caption{Cross-section of the total potential along the $y$-axis at $x = x_0=257.4$\,nm, which is taken at one of the surface optical lattice's minimums at $z = 0$. \label{fig7}}
	\end{figure}  
	
Figure\,8 shows the bound states of a trapped atom along the $x$-axis at $y=z=0$, where the trap potential minimum is located. The potential contains several stable bound energy states owing to the low tunneling probability of atoms towards the dielectric surface of the waveguide at $x$ = 0 due to the high optical dipole potential barrier. The potential is characterized by the frequency of harmonic oscillations near the bottom of $f_x=\Omega_x/(2\pi)=54$\,kHz, where the trap minimum is located at $x_0=257.4$\,nm.   
	
	\begin{figure}
	\includegraphics[width=10.5 cm]{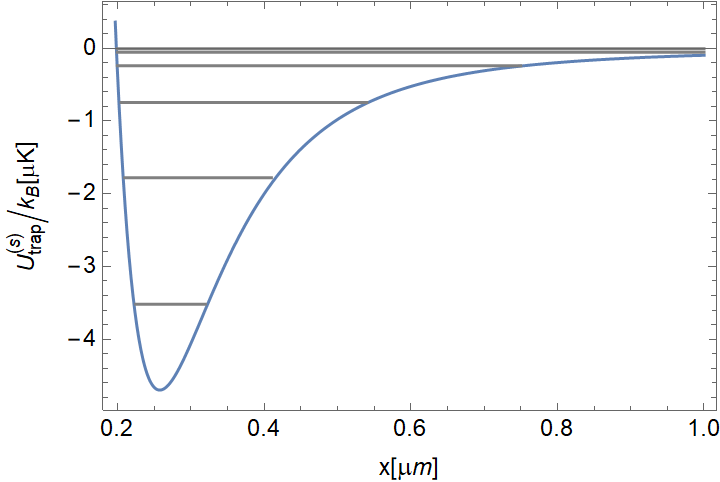}
	\caption{Cross section of the trapping potential along the $x$ axis at $y=z=0$ and its bound states for trapped atoms. \label{fig8}}
	\end{figure}  
	
These calculations demonstrate that the relatively weak trapping potential of the surface optical lattice can sustain only a limited number of bound-energy states for strontium atoms. Ideally, this number should be reduced to one in each dimension, thereby eliminating any uncertainty concerning the position of the atoms relative to the dielectric waveguide surface. This would minimize the nonuniform broadening of the clock transition caused by the van der Waals interaction between the atoms and the surface. There are several ways to collect atoms in a single ground state of the potential. First, the total potential can be reduced so that only a single bound energy state exists in each of the three dimensions. Second, the trapped atoms can be cooled further to condense them in the ground energy state. Third, the depth of the potential could be briefly reduced to clear all but the lowest bound states. 
	
To estimate the spontaneous scattering rate of photons by the trapped atoms at a given trapping magic wavelengths of light, it is sufficient to consider the strongest strontium cooling transition $5s^ 2\,^1S_0 \rightarrow 5s5p\,^1P_1$ at 461\,nm. Using a two-level approximation, the scattering rate for light at 390 nm at $x=x_0$ is $\Gamma_{scat}=0.126$\,s$^{-1}$, and for light at 813\,nm, it is  $\Gamma_{scat}=0.027$\,s$^{-1}$. Therefore, the spontaneous scattering of the trapping light should not prevent the clocks from operating, even over long interrogation times.
		
\subsection{Clock potential for 390 nm trapping light}
		
A configuration of a single, blue-frequency-detuned surface optical lattice is shown in Fig.\,2. This includes a superposition of a blue-detuned evanescent-field lattice and an additional repulsive background potential, which is formed by two counter-propagating TE$_{00}$ modes at the magic wavelength of 390\,nm, with different amplitudes. Transverse trapping of the atoms is provided by the TE$_{01}$ mode, whose amplitude is greater than the TE$_{00}$ mode intensity at the optical lattice minima.
	
In this configuration, the trapped atoms are attracted solely by the van der Waals forces to the surface of the waveguide. The advantage of this approach is that red-frequency-detuned light at 813\,nm, along with the associated fluctuations in the bottom position of the trap caused by the intensity of that light, are both excluded. Another advantage is the small spatial period of the blue optical lattice at 390\,nm:  $\lambda_b/2/n_{eff}=169.2$\,nm. This enables more atoms to be trapped within the same waveguide length. However, at large distances from the waveguide dielectric surface, the van der Waals force becomes negligible. This means that it is impossible to trap atoms at these distances without an attractive optical potential from another red-detuned evanescent-light field. Consequently, this configuration inherently necessitates a stronger van der Waals interaction between the stored atoms and the waveguide surface to achieve the same total light-field power.
	
For example, the minimum of the blue lattice is located at $x_0=214.6$\,nm, where the total potential depth is equal to 4.76\,$\mu$K, which corresponds to the potential depth of the trapping potential along the $x$-axis, for blue modes intensities of $I_{b,TE_{00}^+}(0,0)=4\times10^{10}$\,W/m², $I_{b,TE_{00}^-}(0,0)=4\times10^{9}$\,W/m² and $I_{b,TE_{01}}(0,y_m)=2\times10^{10}$\,W/m². Here $TE_{00}^+$ and $TE_{00}^-$ are counter-propagating main modes at 390\,nm. The lateral depth of the optical lattice sites is 1.52\,$\mu$K, and the depth of the optical lattice along the $z$-axis is 2.77\,$\mu$K, corresponding to $2.88 E_{rec}^{lat}$, where $E_{rec}^{lat}=h^2 n_{eff}^2/(2M \lambda_b^2)$. The corresponding hopping rate of atoms between neighboring sites of the optical lattice is therefore estimated to be $J/h \sim 3340$\,Hz. The Lamb-Dicke parameter of the lattice is $\eta=0.26$. 
 
 	\begin{figure}
 	\includegraphics[width=9.0 cm]{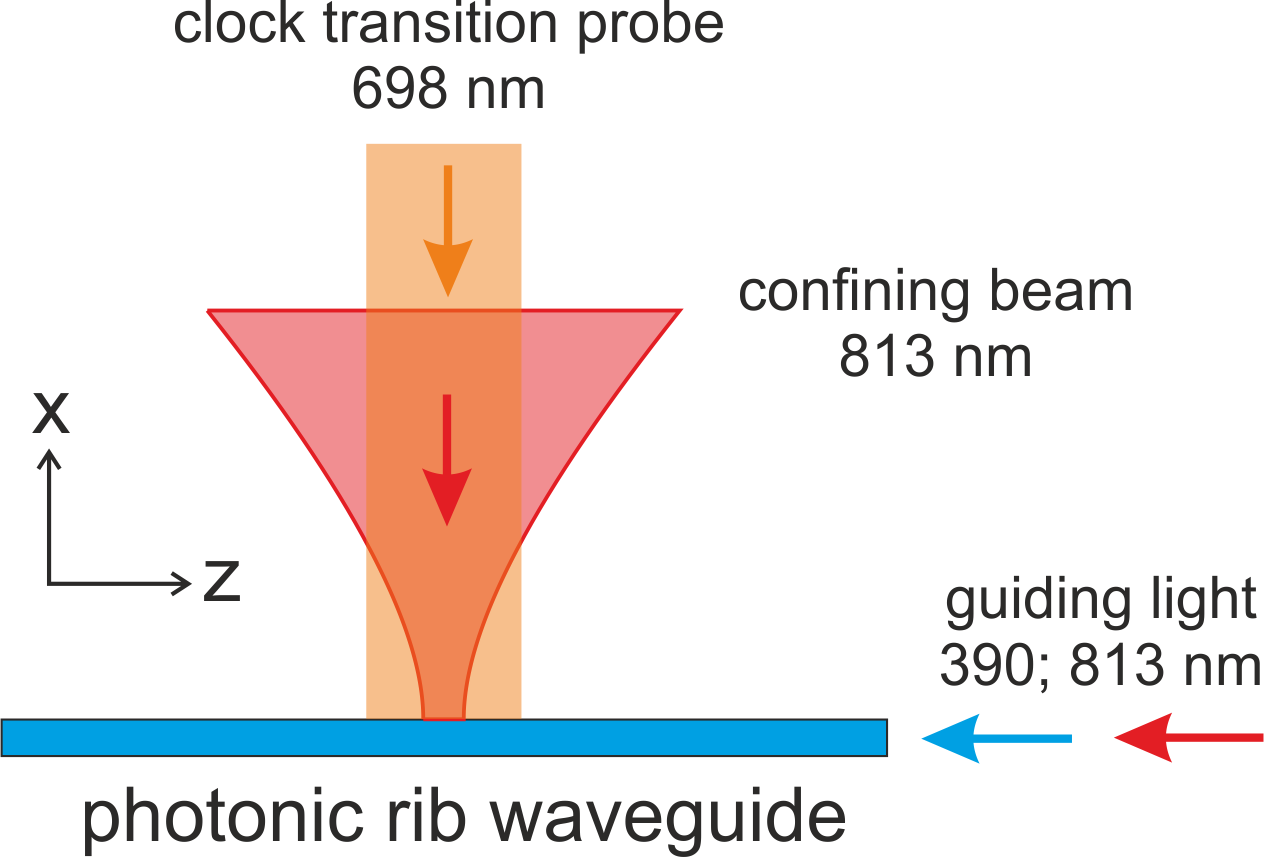}
 	\caption{Configuration of additional laser beams is required for the longitudinal confinement of atoms in the waveguide and for interrogating the clock transition along the $x$-axis. \label{fig9}}
 	\end{figure} 
 
This relatively high hopping rate should result in the trapped atoms diffusing spatially along the optical lattice. However, this diffusion should primarily affect the non-uniform spectral broadening of the clock transition without directly impacting the clocks' accuracy. This diffusion can be limited in several ways. Firstly, additional repulsive potential barriers can be introduced on either side of the optical lattice to prevent the atoms from expanding beyond them. Secondly, an additional attractive, transverse, focused laser beam at the 'magic' wavelength of 813\,nm can be introduced (see Fig.\,9), which would provide an additional trapping potential along the waveguide ($z$-axis). Provided the waist of this laser beam coincides with the minimum of the trapping potential, this should not alter the position of the trap minimum with respect to the waveguide surface. Therefore, power fluctuations in the beam should not alter the frequency of the clock transition. An additional advantage of introducing an additional longitudinal Gaussian potential is the suppression of the tunnelling rate of atoms between adjacent sites of the optical lattice due to a relative shift in their potentials.
 
Compared to free-space atomic optical clocks, waveguide-based clocks offer the advantage of not requiring an optical lattice. Figure\,9 shows that the clock transition for atoms trapped in the waveguide can be interrogated using a 698\,nm laser beam which is perpendicular to the waveguide surface (along the $x$-axis). As long as the trapped atoms are localized along the x-axis within the sub-wavelength region of space, efficient Lamb-Dicke narrowing of the interrogated optical clock transition is expected. 

According to the concept presented in Fig.\,9, it is possible to trap atoms along the waveguide with an additional red-detuned laser beam at the 'magic' wavelength of 813\,nm without any optical lattice component, using blue-detuned light at $390$\,nm. In this configuration, trapping along the $z$-axis is provided by the red-detuned, transverse-focused beam; trapping along the $y$-axis is provided by the TE$_{00}$ and TE$_{01}$ blue-detuned modes; and trapping along the $x$-axis is provided by a balance between the attractive van der Waals force and the repulsive force of the blue-detuned modes. 

For example, for intensities of the TE$_{00}$ and TE$_{01}$ modes at $390$\,nm of $I_{b,TE_{00}}(0,0)=3.6\times10^{10}$\,W/m$^2$ and $I_{b,TE_{01}}(0,y_m)=4.0\times10^{10}$\,W/m$^2$, the minimum of the total trapping potential depth of $2.8$\,$\mu$K is located at $x_0=271.7$\,nm. The oscillation frequency of atoms near the bottom of the potential along the $x$-axis is equal to $f_x=41.4$\,kHz. In a case of interrogation of the trapped atoms along this axis, the corresponding Lamb-Dicke parameter is equal to $\eta=0.34$. The lateral depth of the potential in the $x0y$-plane equalling $0.71$\,$\mu$K with oscillation frequency of $f_y=9.4$\,kHz. The corresponding lateral tunneling rate of the atoms from the ground vibration state of the trap is equal to $\Gamma_{tun}(n=0)=1$\,s$^{-1}$. An additional increase in the lateral potential depth is expected due to the red-detuned, transverse-focused laser beam at the 'magic' wavelength of $813$\,nm. This configuration requires $14.3$\,mW of power in the TE$_{00}$ mode and $17.8$\,mW in the TE$_{01}$ mode.

The potential benefits of the last approach in terms of the accuracy of AOAC strontium clocks are discussed in subsection B of Section IV.

\section{Expected performance of AOAC clocks in different configurations}

\subsection{Case of 390 nm and 813 nm magic wavelengths}
	
For the trapping of strontium atoms in a two-color evanescent light trap with magic wavelengths (390 and 813 nm), the van der Waals interaction of the atoms with the dielectric surface of the photonic waveguide is the only cause of the frequency shift of the clock transition. The difference in the van der Waals potential for atoms in two different clock states results in a constant offset in the clock frequency at the minimum of the trapping potential, $x=x_0$.  
	
	\begin{equation}  \label{EQ2} 
	\Delta f_{clock}(x_0)\approx -\frac{1}{h} \frac{\epsilon-1}{\epsilon+1} \frac{C_3^{p}-C_3^{s}}{x_0^3},				
	\end{equation} 
where $C_3^{(s)}$ = 2.190 $\times 10^{-48}$\,Jm$^3$ and $C_3^{(p)}$ = 2.352 $\times 10^{-48}$\,Jm$^3$ are the van der Waals coefficients for the $5^1S_0$ and $5^3P_0$ clock states of the strontium atom \cite{derevianko2010electric}, $\epsilon=3.9$ is the electric permittivity of silicon dioxide, and $h$ is the Planck constant. For the previously considered maximum surface light intensities of $I_{r,TE_{00}}^{\pm}(0,0)=1\times10^{9}$\,W/m$^2$ and $I_{b,TE_{00}}(0,0)=4\times10^{10}$\,W/m$^2$, $I_{b,TE_{01}}(0,y_m)=2\times10^{10}$\,W/m$^2$, the corresponding frequency shift of the clock transition at $x_0=257.4$\,nm is $\Delta f_{clock}(x_0)=-8484.45$\,Hz (Fig.\,10). This corresponds to a clock fractional frequency accuracy of $2.0 \times 10^{-11}$, in agreement with the predictions in ref. \cite{derevianko2010electric}.
	
	\begin{figure}
	\includegraphics[width=10.5 cm]{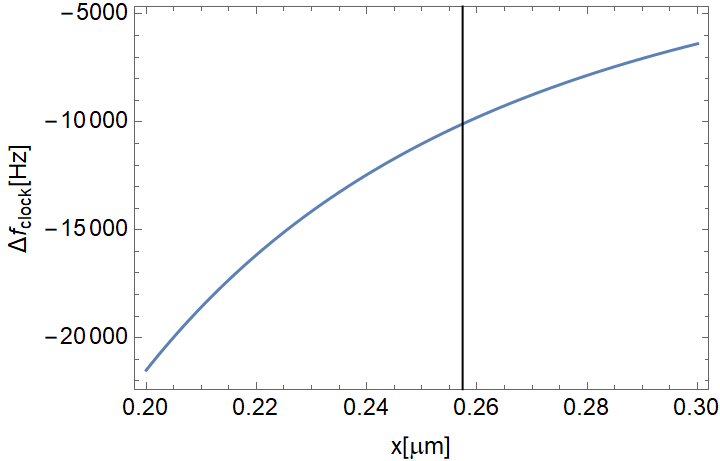}
	\caption{Frequency shift of the strontium clock transition of due to van der Waals interaction with the photonic waveguide's dielectic surface is shown as a function of its distance from the surface. The vertical line shows the position of the trap minimum at $x_0=257.4$\,nm. \label{fig10}}
	\end{figure} 
	
A short-term frequency stability of the clock can be estimated according to the formula \cite{ludlow2015optical} 
	
	\begin{equation}  \label{EQ3} 
	\sigma_y(\tau) = K \frac{\Delta f_{clock}^*}{f_{clock}} \frac{1}{\text{SNR}} \sqrt{\frac{T_c}{\tau}},				
	\end{equation} 
where $\Delta f_{clock}^*$ is determined not only by the natural linewidth of the clock transition but also by other factors, the SNR (signal-to-noise ratio) of the detected atoms is given by $T_c$ (the cycle time of the clock) and $\tau$ (the averaging time) and $K$ is a parameter which depends on the line shape. For interrogation of the clock transition by square Rabi pulses, $K \approx 0.6$. One of the main factors contributing to the non-uniform broadening of the clock transition is a decrease in the intensity of the propagating waveguide modes owing to propagation losses. Changes in the intensity of the trapped evanescent light waves result in changes in the position of the minimum trapping potential relative to the waveguide surface, leading to a corresponding change in the van der Waals interaction. For a 50\,$\mu$m-long suspended silica waveguide with propagation losses of 0.385 dB/cm \cite{chen2012fabrication}, the expected broadening of the clock transition is expected to be $\frac{\Delta f_{clock}^*}{f_{clock}}=9\times 10^{-15}$. In the event of fractional intensity instability of the trapping optical fields of $10^{-3}$, the signal-to-noise ratio in the clock is dominated by the quantum projection noise, which is proportional to $\sqrt{N}$, where $N$ is the total number of trapped atoms in the surface optical lattice. For a surface optical lattice length of 50\,$\mu$m and a period of 353 nm, the total number of stored atoms with a population of one per site is $N=141$. This gives the QPN a limited short-term stability of the clock at $T_c=1$\,s, assuming absence of dead time of the clock and no technical noise, of approximately $\sigma_y(\tau) \approx 4.5\times 10^{-16}/ \sqrt{\tau}$. Reducing the length of the surface optical lattice to 10 $\mu$m reduces the broadened relative linewidth of the clock transition to  $\frac{\Delta f_{clock}^*}{f_{clock}}=1.8\times 10^{-15}$. With $N=28$, this gives a clock short-term frequency stability of $\sigma_y(\tau) \approx 2.0\times 10^{-16}/ \sqrt{\tau}$.
	
In this case, the long-term stability of the clock is mainly determined by the long-term stability of the trapped optical fields. In the event of linear drift in the intensities of the trapping light fields, the clock's long-term stability is determined by the formula
	
	\begin{equation}  \label{EQ4} 
		\sigma_y^{drift}(\tau) \approx  \frac{b \tau}{f_{clock} \sqrt{2}} ,			
	\end{equation} 
where $b=df_{clock}/dt$ is the rate of change in the clock frequency.	According to calculations, a long-term drift of the relative intensities of the 390\,nm and 813\,nm modes by $10^{-3}$ should lead to long-term stability of the clocks on the corresponding time scale to that of the intensity change, which is $2 \times 10^{-14}$.

\subsection{Compensation of van der Waals frequency shift with an optical AC Stark shift}

The negative van der Waals frequency shift of the atomic clock transition at the minimum of the trapping potential ($x_0$) can be offset for by the positive optical AC Stark shift of the clock transition. This shift can be produced by either the blue-detuned ($\lambda_b$) or the red-detuned ($\lambda_r$) evanescent light fields of the corresponding frequency.

Initially, it is assumed that only one of the frequency components ($\lambda_b$ or $\lambda_r$) of the trapping fields has a non-magic frequency, thereby producing an intensity-dependent optical AC Stark shift of the clock transition.

Complete suppression of the van der Waals frequency shift at $x=x_0$ occurs when the following condition is met:

	\begin{equation}  \label{EQ5} 
	\Delta U_{b,r}^{*} = \frac{C_3^{(p)}-C_3^{(s)}}{x_0^3 e^{-2 x_0/d_{b,r}}} ,			
	\end{equation}
where $\Delta U_{b,r}^{*}=U_{b,r,\text{TE}_{00}}^{(p)}(x=0,y=0)-U_{b,r,\text{TE}_{00}}^{(s)}(x=0,y=0)$ is the light-induced change in the energy of the clock transition at the surface of the waveguide. 

	\begin{figure}
	\includegraphics[width=10.5 cm]{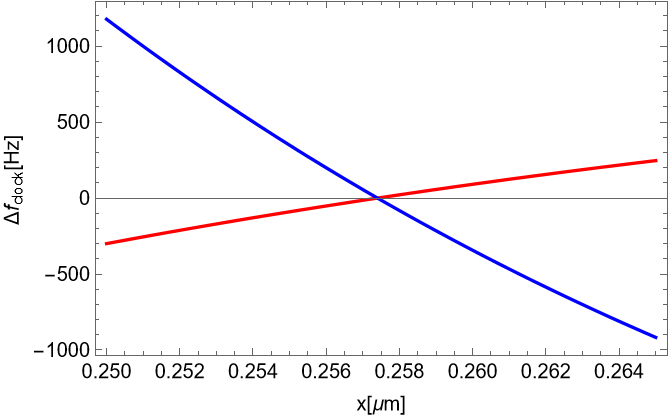}
	\caption{The frequency shift of the strontium atom clock transition is due to the van der Waals interaction with the photonic waveguide's dielectric surface. This is compensated for at the minimum of the trapping potential ($x = x_0 = 257.4$\,nm) by either the optical AC Stark shift of the 'non-magic' blue-detuned $390.5$\,nm mode (blue line) or the optical AC Stark shift of the 'non-magic' red-detuned $933$\,nm mode (red line). \label{fig11}}
\end{figure} 	

First, we make the assumption that the red-detuned frequency component of the trapping light is "magic" with a wavelength of $\lambda_r=812.9$\,nm, while the blue-detuned component is "non-magic" with a wavelength of $\lambda_b=390.5$\,nm. For $x_0=257.4$\,nm, the condition (5) gives us a value of $\Delta U_{b}^{*}/h=1.44\times10^7 $\,Hz. According to the calculated polarizabilities of the $5^3P_0$ and $5^1S_0$ states (see Fig.\,3), the above conditions correspond to a wavelength of $\lambda_b=390.5$\,nm for the blue-frequency-detuned mode. In order to maintain the minimum of the trap at $x_0=257.4$\,nm, the intensity of the red-detuned light must be $I_{r,TE_{00}}(0,0)=3.59\times10^{8}$\,W/m$^2$, while the intensity of the blue-detuned light is $I_{b,TE_{00}}(0,0)=4.0\times10^{10}$\,W/m$^2$. The dependence of the frequency shift of the clock transition on $x$ is depicted in Fig.\,11 as a blue line.

Another option is to compensate for the van der Waals frequency shift of the strontium clock transition by using a 'non-magic' red-detuned frequency component of the trapping light at $\lambda_r = 933$\,nm, while the blue-detuned component is a 'magic' frequency at $\lambda_b = 389.9$\,nm. For the minimum of the trapping potential is located at $x_0=257.4$\,nm and the above-considered geometry of the suspended optical rib waveguide made of silica, the penetration depth of the evanescent light field for the $TE_{00}$ mode at $\lambda_r = 933$\,nm is $d_r = 271$\,nm. This results in van der Waals compensation at $U_r^{*}/h = 5.671 \times 10^4$\,Hz. In this case, the corresponding intensities of the two trapping modes are $I_{r,TE_{00}}(0,0) = 3.65 \times 10^8$ W/m$^2$ and $I_{b,TE_{00}}(0,0) = 4.03 \times 10^{10}$\,W/m$^2$. The dependence of the frequency shift of the clock transition on $x$ is depicted in Fig.\,11 as a red line.
In both cases above, the potential depth along the x-axis is approximately $5.5$\,$\mu$K.

The curves in Fig.\,11 show how the frequency shift of the strontium clock transition varies with distance $x$, for constant powers of the two trapping optical modes. Any change in the mode intensities has two different effects. Firstly, a change in the intensity of the 'non-magic' mode directly affects the frequency shift of the transition via the optical AC Stark shift. This manifests as an upwards or downwards shift in the corresponding curve in Fig.\,11. Secondly, a change in the intensity of either of the two modes shifts the trap minimum $x_0$. This also leads to a corresponding frequency shift of the clock transition due to a different van der Waals potential at the new location. For the blue-detuned 'non-magic' mode, it turns out that the two aforementioned frequency shift mechanisms have opposite signs, which essentially reduces the total power-dependent frequency shift.
For the 'magic' wavelength mode, only the power-dependent frequency shift of the clock transition is related to the shift of the minimum of the trapping potential.

	\begin{figure}
	\includegraphics[width=10.5 cm]{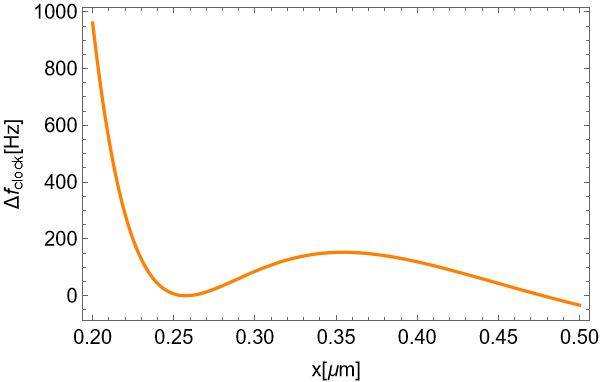}
	\caption{The frequency shift of the strontium atom clock transition due to van der Waals interaction with the photonic waveguide's dielectric surface is compensated for at the minimum of the trapping potential ($x = x_0 = 257.4$ nm) by the optical AC Stark shift of the 'non-magic' blue-detuned $390.07$\,nm mode and 'non-magic' red-detuned $933$\,nm mode. The wavelengths of the two frequency components of the trapping light are chosen to compensate for the absolute van der Waals frequency shift of the strontium clock transition at the minimum of the trapping potential and to ensure that the spatial derivative of the total frequency shift around that point is zero.  \label{fig11}}
\end{figure} 

In the above example, increasing the intensity of the $390.5$\,nm field by a relative fraction of $10^{-3}$ results in an $8.48$\,Hz frequency increase for the clock transition due to the optical AC Stark shift, given the mode intensities. However, due to the change in the position of the potential minimum at the increased intensity of the blue-detuned trapping optical field, the total frequency shift is $1.27$\,Hz, giving a suppression factor of $6.7$. The frequency shift of the clock transition due to a relative change of $10^{-3}$ in the intensity of the red-detuned mode at $812.9$\,nm is $1.94$\,Hz. Assuming the intensity fluctuations of the two modes are independent, the clock's accuracy is $5.4\times10^{-15}$.  The suppression of the frequency dependence of the clock transition in this configuration should also suppress the short-term QPN-limited stability of the clock with an optical lattice length of $50$\,$\mu$m (see Subsection A) to $\sigma_y(\tau) \approx 4.5\times 10^{-16}/ \sqrt{\tau}$.  

As described above, in the case of 'non-magic' red-detuned and 'magic' blue-detuned trapping optical fields, the mechanism for compensating for the optical AC Stark frequency shift by shifting the position of the trap minimum does not work. In the presence of intensity fluctuations at the $10^{-3}$ level, the expected accuracy of such clocks is $1.9\times10^{-14}$.

Setting both trapping light components to 'non-magic' values fully compensates for the van der Waals force at the minimum of the trapping potential, setting the potential derivative to zero at this point. Figure 12 shows the compensated van der Waals frequency shift of the strontium clock transition for wavelengths of the trapping modes of $390.07$\,nm and $933$\,nm. The other parameters are: $x_0 = 257.4$\,nm, $U_r^*/h = 4.495 \times 10^4$\,Hz, $U_b^*/h = 2.373 \times 10^6$\,Hz, $I_{r,TE_{00}}(0,0) = 1.61 \times 10^8$ W/m$^2$, $I_{b,TE_{00}}(0,0) = 3.41 \times 10^{10}$ W/m$^2$. The total potential depth along the $x$-axis is $4.2$\,$\mu$K.
Although the frequency shift related to the different distances of the trapped atoms from the waveguide surface is essentially suppressed, the optical AC Stark shift of the clock transition frequency is not, since both modes operate at 'non-magic' wavelengths. Consequently, the accuracy of such clocks for a $10^3$ fractional intensity stability in both optical modes is approximately $1.6\times10^{-14}$. 

Another approach, which uses only the blue-detuned TE$_{00}$ and TE$_{01}$ modes of the waveguide at a wavelength of $\approx 390.5$\,nm and a focused, transverse confining laser beam at $813$\,nm, as shown in Fig.\,9, can provide even greater clock accuracy. For the trapping potential parameters described in Subsection C of the previous section, the clock's accuracy within the power drift of the blue-detuned trapping optical fields is expected to be $3.9\times10^{-15}$. This improvement is mainly due to the absence of the red-detuned trapping mode and the intensity fluctuations associated with it.    

In conclusion, the optimal configuration for compensating for the van der Waals shift in atomic optical clock transitions in AOAC clocks involves using an optical AC Stark shift produced by a 'non-magic' blue-detuned photonic waveguide mode, while the red-detuned mode is either a 'magic' mode or absent.    

\section{Outlook}

It has been shown that optical strontium clock based on all-optical chips can potentially demonstrate a short-term frequency stabilities at $\sigma_y(\tau) \approx 2.0\times 10^{-16}/ \sqrt{\tau}$. For a linear drift in the relative intensity of the trapping light fields of up to $10^{-3}$, the long-term frequency stability in this configuration is expected to be approximately $2.9\times10^{-15}$. 
In addition, the van der Waals potential at the minimum of the trapping potential can be essentially compensated by the differential optical AC Stark shift of the clock transition. This is provided that the wavelength of the blue-detuned optical field is greater than the 'magic' wavelength, while the red-detuned trapping light is at 'magic' wavelength. The clock's accuracy could be approximately $5\times10^{-15}$ if the relative intensity of the trapping light fields remains stable over the long term by a factor of $10^{-3}$. If the relative trapping field intensities remain stable over the long term by a factor of $10^{-4}$, the long-term frequency stability and accuracy of the clock are expected to improve by a factor of ten.

Although this study only considered strontium atoms, the results can also be applied to other alkali earth elements, such as calcium and ytterbium.

Thulium is one of the most promising atoms for all-optical atom chip clocks. This atom is already used in standard optical lattice clocks \cite{golovizin2019inner}. Its inner-shell magnetic dipole (M1) clock transition between the fine-structure components of the ground electronic state, $|4f^{13}6s^2,J=7/2,F=4,m_F=0 \rangle \rightarrow |4f^{13}6s^2. J=5/2, F=3, m_F=0 \rangle$ at 1.14\,$\mu$m, exhibits an exceptionally small differential scalar polarizability, which is the main advantage of this atom. This leads to a very small differential van der Waals coefficient between the two clock states of approximately $10^{-4} - 10^{-5}$. It was found that thulium has a red-detuned magic wavelength at 813.32\,nm \cite{golovizin2019inner}. Although it is clear that the blue-detuned magic wavelength for thulium atoms should be searched for on either side of the laser cooling transitions at 410\,nm and 506\,nm, or around 390\,nm as with strontium atoms, it has not yet been identified. All-optical thulium atom chip clocks could potentially achieve an accuracy of $10^{-17}$ to $10^{-18}$, taking into account the small van der Waals differential and suppressed black-body frequency shift of the thulium optical clock transition.

Erbium is another rare-earth metal found in the lanthanide series of the periodic table, which is very similar to thulium. The proposed optical lattice atomic clocks for erbium atoms have a wavelength of the clock transition of 1299\,nm and the red-detuned magic wavelength of about 807\,nm \cite{PhysRevA.88.032509, patscheider2021observation}.

The accuracy of these clocks depends on the accuracy of the intensities of the trapping laser fields. One possible solution is to calibrate the clocks periodically by measuring the trapping potential depth directly. One method of achieving this would be to measure the exact energies of the bound states of the potential along the x-axis.

One of the most promising applications of 'magic' surface optical lattices, which is considered in this paper, is active atomic clocks \cite{matusko2024superradiant, dubey2025modeling}. These clocks exploit the superradiance of ultra-cold atoms in an optical resonator, which may be an optical cavity \cite{norcia2016superradiance}, a photonic crystal waveguide \cite{goban2015superradiance}, a hollow optical fiber \cite{okaba2019superradiance} or a photonic ring resonator \cite{zhou2024trapped, margalit2026single}. Notably, both Ref. \cite{norcia2016superradiance} and Ref. \cite{okaba2019superradiance} have demonstrated superradiance in strontium atoms by exploiting the small linewidth of spin-forbidden electric dipole transitions between singlet and triplet electronic states.
In the case of planar all-optical atom chips, ring photonic resonators can be used to create integrated active optical atomic clocks. Trapping corresponding atoms in a 'magic' circular optical lattice within a ring photonic resonator suppresses most frequency shifts and non-uniform broadening of the optical clock transition. However, shifts caused by van der Waals interactions between the atoms and the dielectric surface of the waveguide remain. Consequently, the frequency stability and accuracy of the clocks are determined by these shifts. This surface optical lattice enables strong dipole–dipole interactions between the trapped atoms and strong coupling to the resonator's single pumping mode at the frequency of the strontium clock transition. In principle, such clocks can emit a continuous superradiant highly coherent light with an extremely small linewidth at the frequency of the atomic clock transition.

\begin{acknowledgments}
The author acknowledges the support of the National Physical Laboratory for the preparation of this manuscript.
I thank Sean Mulholland, Ian Hill, and Geoffrey Barwood for their comments and corrections on the paper.
\end{acknowledgments}

\nocite{*}
\bibliography{Ref2026clock}
\end{document}